\documentclass[aps,twocolumn,showpacs,floats,prl]{revtex4}
\usepackage{graphicx}
\usepackage{dcolumn}
\usepackage{bm}
\usepackage{color}
\usepackage{subfigure}
\usepackage{amsmath}
\usepackage{amsmath,amsfonts,amssymb,graphicx}
\usepackage{braket}

\begin{document}

\author{L. Parisi and S. Giorgini}
\affiliation{Dipartimento di Fisica, Universit\`a di Trento and CNR-INO BEC Center, I-38123 Povo, Trento, Italy}

\title{Quantum droplets in one-dimensional Bose mixtures: a quantum Monte-Carlo study}
\begin{abstract} 
We use exact Quantum Monte Carlo techniques to study the properties of quantum droplets  in two-component bosonic mixtures with contact interactions in one spatial dimension. We systematically study the surface tension, the density profile and the breathing mode as a function of the number of particles in the droplet and of the ratio of coupling strengths between intra-species repulsion and inter-species attraction. We find that deviations from the predictions of the generalized Gross-Pitaevskii equation are small in most cases of interest.
\end{abstract}
\pacs{05.30.Fk, 03.75.Hh, 03.75.Ss} 
\maketitle

\section{I. Introduction}
One of the recent progress in ultracold atoms is the observation of liquid quantum droplets which were first observed in dipolar gases~\cite{Rosensweig-instability,PhysRevLett.116.215301,Ferrier_Barbut_2016,Schmitt_2016,PhysRevX.6.041039,PhysRevLett.122.130405} and later in Bose mixtures~\cite{Cabrera301,PhysRevLett.120.135301,PhysRevLett.120.235301,PhysRevLett.122.090401,PhysRevResearch.1.033155}. In the case of dipolar atoms, the discovery initially came as a surprise because the stable droplets observed in the experiment where not compatible with mean-field theories which predicted a collapse of the system. It was later realized that repulsive quantum fluctuations contrast the mean-field attraction and stabilize the system in a liquid-like phase~\cite{PhysRevX.6.041039}. In nature liquids can arise as a result of a competition between attraction and repulsion. In dipolar gases the repulsion is provided by the short-range interaction while the attraction is provided by the long-range dipolar force. However, the presence of a long-range interaction is not essential to the droplet formation mechanism. For instance, a similar mechanism arise in bosonic mixtures subject to contact forces which provide repulsive interactions between particles of the same species and attractive interactions between particles of different species. In this case the formation of quantum droplets was first predicted theoretically~\cite{PhysRevLett.115.155302} and then confirmed experimentally~\cite{Cabrera301}.

In three dimensions (3D) uniform mixtures are predicted to collapse, according to mean-field theory, when the inter-species attraction exceeds the intra-species repulsion. However, when quantum fluctuations are included in the theory, their repulsive effect can balance the mean-field attraction and the system becomes stable. Moreover, the energy acquires a minimum at a non-zero density, which is the hallmark of a liquid state at zero temperature. Furthermore, it was shown that liquid states can arise also in lower dimensions~\cite{PhysRevLett.117.100401} and, in particular, in one-dimensional (1D) configurations~\cite{parisi-liquid-mixture}. In 1D the problem is enriched compared to the 3D case by the enhanced role of quantum fluctuations~\cite{giamarchi1DBook,parisi-liquid-mixture}. Mixtures in 1D can be realized experimentally by loading the atoms in elongated tubes confined by a two-dimensional optical lattice~\cite{Nagerl1} or using micro-traps~\cite{ViennaCooling}. In contrast to 3D, where it is difficult experimentally to reach strongly correlated regimes because of three-body losses, in 1D such losses are greatly reduced~\cite{Astra2006LBCorr}. Furthermore, the effective 1D scattering length can be tuned from the weakly to the strongly interacting regime, up to the limit of infinite repulsion, also called Tonks-Girardeau gas~\cite{Paredes04,Kinoshita1125,Haller1224}.

Since quantum fluctuations play an important role, it is crucial to rely on theoretical approaches able to describe beyond mean-field corrections. For homogeneous systems in the thermodynamic limit, the Bogoliubov theory includes perturbatively fluctuations at the lowest order and provides an adequate description in the weakly interacting limit~\cite{parisi-liquid-mixture}. However, experiments deal with non-uniform droplets formed by surface effects and containing a finite number $N$ of particles. In the few particle limit one expects to form a small self-bound cluster, whereas for large $N$ one expects to form a droplet with a mostly flat density profile except at the boundaries, which reminds of the shape of a water puddle. To describe static and dynamic properties of these finite-size droplets many theoretical studies use a generalized Gross-Pitaeskii equation (GGP) which includes quantum fluctuations within a local density approximation~\cite{PhysRevLett.115.155302,PhysRevLett.117.100401,PhysRevA.98.013631,PhysRevA.97.053623,PhysRevA.98.051603,PhysRevA.98.013612}. However, the validity of this approach is not clear, even in the weakly interacting limit. To assert the validity of the theory it is therefore important to benchmark its predictions against microscopic non-perturbative theories.

The ground state of the microscopic Hamiltonian can be investigated numerically using exact Quantum Monte Carlo (QMC) techniques~\cite{becca_sorella_2017}. These methods were used to study the properties of the liquid state in 3D Bose mixtures, both in the case of uniform systems and non-uniform quantum droplets~\cite{PhysRevA.99.023618,PhysRevB.97.140502}. In 1D, the few-body problem was investigated in Ref.~\cite{PhysRevA.97.061605} and uniform mixtures in the thermodynamic limit were investigated using QMC techniques in Ref.~\cite{parisi-liquid-mixture}. In particular, it was found that the ground state of a uniform system is a liquid, i.e. the energy displays a minimum at a non-zero density, above a critical ratio $r\approx 0.45(3)$ between attractive and repulsive coupling constants. Below this critical ratio the minimum in the energy corresponds to zero density and the ground state is a gas. Above the critical ratio, a finite system is expected to form self-bound clusters. So far, however, no exact results are available in the literature for these non-uniform droplets.

In this article we study the properties of 1D non-uniform quantum droplets at zero temperature using exact QMC techniques. We calculate the surface energy, the density profiles and the frequency of the breathing mode and we systematically compare our results with the predictions of GGP theory. We find that the GGP equation describes quite well our results, even at small particle numbers and small ratio between attraction and repulsion strength, in regimes where the theory cannot be a priori justified.

The structure of the paper is as follows: In Sec.~II we introduce the main theoretical ingredients of our analysis which include the Hamiltonian defining the model of 1D Bose mixtures, the GGP approach for the homogeneous liquid phase and for the droplet, the many-body droplet model and the sum-rule approach used to extract the frequency of the breathing collective mode. In Sec.~III we briefly discuss the QMC numerical technique employed to calculate the ground-state properties of the liquid droplets. Finally, in Sec.~IV, we present our results for the surface energy, the droplet density profiles as a function of the number $N$ of particles and the frequency of the breathing mode. We also provide a quantitative comparison with the predictions of the GGP theory. Sec.~V contains our concluding remarks.

\section{II. Model}

\paragraph{ { \bf Hamiltonian:}}
In this work we consider a two-component mixture of bosons at zero temperature with contact interactions. The Hamiltonian of the system is given by
\begin{eqnarray}
  H&=&-\frac{\hbar^2}{2m_1}\sum_{i=1} ^{N_1}\frac{\partial^2}{\partial x_i^2} 
  -\frac{\hbar^2}{2m_2}\sum_{\alpha=1}^{N_2} \frac{\partial^2}{\partial x_\alpha^2} + g_{11} \sum _{i<j} \delta(x_i - x_j) \nonumber\\
  &+& g_{22} \sum_{\alpha < \beta}\delta(x_\alpha - x_\beta) + g_{12} \sum_{i,\alpha} \delta(x_i-x_\alpha) \;,
\label{Hamilton}
\end{eqnarray}
where $m_1$ and $m_2$ are the masses of the bosons respectively of the first and second component, $g_{11}$ and $g_{22}$ are the coupling constants for the intra-species two-body interaction between atoms of the same component and $g_{12}$ is the inter-species two-body coupling constant. Furthermore, the labels $i, j$ and $\alpha, \beta$ refer respectively to the coordinates of particles belonging to the first and second component. For the sake of simplicity we only consider symmetric mixtures with equal masses $m_1=m_2=m$, equal  intra-species coupling constants $g_{11}=g_{22}=g$ and equal particle numbers $N_1=N_2=N/2$. Moreover, we will consider repulsive intra-species interactions with $g>0$ to ensure the stability of the system and an attractive inter-species interaction with $g_{12}<0$.

\paragraph{ \bf{Generalized Gross-Pitaevskii equation:} }
Here we review the GGP approach to study bosonic mixtures in 1D, as described in Refs.~\cite{PhysRevLett.117.100401,PhysRevA.98.013631}. For a uniform mixture in the weakly interacting regime one can apply the Bogoliubov theory which yields the following energy per unit volume~\cite{PhysRevLett.117.100401}
\begin{eqnarray}
  \label{BogoliubovEnergy}
  \frac{E}{V}&=& \frac{(g - |g_{12}|)}{4}n^2 \nonumber\\
  &-&\frac{\sqrt{m}n^{3/2}}{3\sqrt{2}\pi\hbar}\left[ (g + |g_{12}|)^{3/2}  + (g - |g_{12}|)^{3/2} \right] \;.
\end{eqnarray}

The first term in Eq.~(\ref{BogoliubovEnergy}) is the mean-field interaction term containing the total density $n=n_1+n_2$. The second term arises instead from quantum fluctuations treated at the lowest order, also called Lee-Huang-Yang (LHY) corrections. The above energy functional is applicable in the high density limit $n|a| \gg 1$, where $a=-\frac{2\hbar^2}{mg}$ is the 1D scattering length between atoms. Notice that this behavior is different from the 3D case where Bogoliubov theory is applicable only in the opposite limit of very low density. 
Notice also that the LHY corrections only make sense if $|g_{12}|< g$ and they are negative. As a consequence the compressibility becomes negative and the system collapses when $|g_{12}|> g$. Instead, when $|g_{12}|< g$ but $|g_{12}|$ is close to $g$, the attractive LHY correction can balance the mean-field repulsion. Crucially the mean-field and the quantum fluctuation term scale with different powers of the density and the energy functional acquires a minimum at the equilibrium density $n_{eq}$ given by
\begin{equation}
  \label{n_eq_bg}
  n_{eq}=\frac{2m}{\hbar^2}\left(\frac{   \left[ (g + |g_{12}|)^{3/2}  + (g - |g_{12}|)^{3/2}             \right]  }{3\pi(g-|g_{12}|)}\right) ^2 \;.
\end{equation}

The presence of a finite density minimizing the energy means that the ground state is a liquid. Notice also that $n_{eq}$ increases as the ratio $|g_{12}|/g$ between attraction and repulsion increases and in the limit $|g_{12}|\rightarrow g$ the equilibrium density becomes large. As mentioned above, this regime, corresponding to $n|a|\gg 1$, is the condition required for the validity of the GGP theory. However, in a system with a finite number of atoms, the effect of surface tension leads to a rearrangement of the particles in a self-bound non uniform configuration. To describe such non uniform systems one can use an approach based on density functional theory with the energy given by Eq.~(\ref{BogoliubovEnergy}). This approach yields the following time-dependent equation
\begin{equation}
  \label{ggp}
  \begin{array}{cl}
i\hbar \frac{\partial \psi}{\partial t} = &  -\frac{\hbar^2}{2m} \frac{\partial^2}{\partial x^2}\psi + \frac{1}{2}(g-|g_{12}|)|\psi|^2\psi  \\
    & - \frac{\sqrt{m}}{\hbar\pi 2^{3/2}}  \left[ (g + |g_{12}|)^{3/2}  + (g - |g_{12}|)^{3/2}    \right]   |\psi|\psi  \;,
  \end{array}
\end{equation}
where $\psi$ is the wavefunction of the system whose modulus square is equal to the total density $n$.  We refer to Eq.~(\ref{ggp}) as the GGP equation. The solutions of the above equation were investigated and discussed in detail in Ref.~\cite{PhysRevA.98.013631}.

\paragraph{ \bf{The liquid-droplet model:}}
The liquid-droplet model relates finite-size energy corrections with the surface energy of the droplet. In particular, it predicts that in the limit of large particle numbers the energy per particle in 1D can be written as 
\begin{equation}
  \frac{E}{N}= E_{B} + E_S \frac{1}{N} + ... \;,
\end{equation}
where $E_{B}$ is the bulk energy and the first correction due to a finite atom number scales as $1/N$ with a proportionality coefficient given by the surface energy $E_S$. Higher order corrections are neglected. The bulk energy is just the energy per particle of an homogeneous system in the thermodynamic limit. It is worth mentioning that the liquid-droplet model has proved successful in describing the properties of clusters of liquid $^4$He~\cite{PhysRevLett.50.1676,PhysRevB.45.852}. According to the model the surface tension of the droplet is obtained by dividing the surface energy $E_S$ by the area of the surface. As in 1D the surface is made of only two points one can just divide the surface energy by two to obtain the surface tension: $\tau=E_S/2$.

\paragraph{\bf{Sum rules:}}
Apart from ground-state properties it is also worth investigating the collective modes of a quantum droplet. However, the accurate study of the dynamics of the system is hard in the beyond mean-field regime. In particular, QMC methods are not well suited to this aim because of the well known sign problem. Fortunately, one can use linear response theory and sum rules obeyed by the response function to obtain reliable estimates of excitation energies from the expectation values of appropriate operators in the many-body ground state, which can be sampled using QMC techniques.

For a small perturbation proportional to a given operator $F$ the information about the linear response of the system is provided by the dynamical structure factor. At zero temperature the dynamical structure factor is defined as~\cite{sandrobook}
\begin{equation}
 S_F(\omega)=\sum_n |\Bra{n} F \Ket{0}|^2 \delta (\omega_{n0} - \omega) \;,
\end{equation}
where $\Ket{n}$ represents the $n_{th}$ excited state, $\Ket{0}$ represents the ground state of the system and $\omega_{n0}=(E_n-E_0)/\hbar$ is the excitation frequency between the two states. Many features of the dynamic structure factor are captured by its moments defined as
\begin{equation}
m_p=\hbar^p\int  S_F(\omega) \omega^p d\omega \;,
\end{equation}
where $p$ is the order of the moment. Let us suppose that the operator $F$ can only excite a single mode, i.e. one has $\Bra{n} F \Ket{0}=0$ for every state but one whose excitation energy is $E_n-E_0=\hbar\omega_F$. We also assume that the expectation value in the ground state of the operator $F$ is zero, i.e. $\Bra{0} F \Ket{0}=0$. Under these assumptions, by performing the integration over frequencies for successive moments of the response function one can derive the relationship  $\hbar\omega_F = \frac{m_{p+1}}{m_p}$. In general, the operator $F$ excites more than just one energy mode and the relationship does not hold. However, one can prove that $\frac{m_{p+1}}{m_p}$  always provides an upper bound for the energy of the excited state and one has
\begin{equation}
\hbar \omega_F \leq \frac{m_{p+1}}{m_p} \;.
\end{equation}

While the moments depend on the dynamical structure factor at all frequencies they can actually be calculated without any detailed knowledge of $S_F(\omega)$. In particular, for an hermitian operator $F$ one can derive the following lowest moments~\cite{sandrobook}
\begin{equation}
m_0=\frac{1}{2}\Bra{0} F^2 \Ket{0} \;,
\label{m0}
\end{equation}
and
\begin{equation}
m_1=\frac{1}{2}\Bra{0} [F,[H,F] ]\Ket{0} \;.
\label{m1}
\end{equation}

The above expectation values can be computed using QMC techniques to sample the ground state of the system.  

\paragraph{The breathing mode:}
The breathing mode corresponds to an oscillation of the size of the cloud and is excited by the operator $F=\sum_{i} x_i^2 + \sum_{\alpha} x_\alpha^2$, where one sums over the total number of particles in the two components. Experimentally, this mode can be excited by rapidly switching on and off an harmonic trapping potential. The breathing mode has been extensively investigated in harmonically trapped single-component 1D Bose gases~\cite{menotti02,grigoryBreathingMode} and it is known to depend on many-body properties of the system such as the equation of state. For this choice of the operator $F$, the moments in Eqs.~(\ref{m0})-(\ref{m1}) can be written as
\begin{equation}
m_0=\frac{1}{2}\left(\Braket{F^2} - \Braket{F}^2 \right) \;,
\end{equation}
and
\begin{equation}
m_1=\frac{ \hbar ^2}{m}\Braket{F} \;,
\end{equation}
where $\Braket{...}$ stands for average on the ground state. Therefore, the frequency of the breathing mode can be estimated using the ratio $m_1/m_0$ as
\begin{equation}
\omega_F=\frac{2\hbar}{m}\frac{\Braket{F}}{\Braket{F^2}-\Braket{F}^2} \;.
\label{breathingModeSumRule}
\end{equation}

The above equation links the frequency of the breathing mode to the size of the droplet and its mean square fluctuations on the ground state.

\section{III. Method}
To compute the ground state of the Hamiltonian in Eq.~(\ref{Hamilton}) we use the Diffusion Monte Carlo (DMC) method~\cite{becca_sorella_2017}. This is a stochastic technique which samples the many-body evolution in imaginary time and is able to yield exact results for the ground state of strongly correlated bosons. To improve the efficiency of the algorithm a guiding wavefunction $\Psi(X)$ is used, where $X=\{x_1,x_2,...x_N\}$ indicates the positions in space of the $N$ particles. We choose a guiding wavefunction of the form
\begin{equation}
 \label{jastrowWave}
  \Psi(X)=\prod _{i<j} f(x_i - x_j) \prod _{\alpha < \beta} f(x_{\alpha}-x_{\beta})\prod _{i,\alpha} g(x_i - x_\alpha) \;,
\end{equation}
where the function $f(x)$ describes two-body correlations between particles of the same species, while $g(x)$ describes two-body correlations between particles of different species. We simulate contact interactions by enforcing the Bethe-Peierls boundary conditions. This allows us to map the original problem with two-body contact interactions to a free-particle problem with a constraint on the wavefunction at vanishing inter-particle distance. In particular, the many-body wavefunction must have a discontinuity in the first derivative at zero inter-particle separation. The discontinuity is fixed by the scattering length between the two particles according to the equation
\begin{equation}
  \label{BP_boundary}
  \frac{\partial}{\partial (x_i - x_j )} \Phi(X) =-\frac{1}{a_{i,j}}  \Phi (X)_{x_i=x_j} \;,
\end{equation}
where $a_{i,j}$ is the scattering length between particles $i$ and $j$. The above boundary condition on the ground-state wavefunction  $\Phi(X)$ can be enforced by choosing a guiding wavefunction which satisfies the same condition. For the guiding wavefunction in Eq.~(\ref{jastrowWave}) the Bethe-Peierls boundary conditions imply that the correlators $f$ and $g$ satisfy the relations $ \frac{d}{d x} f(x)_{x=0}=-f(0)/a$ and $\frac{d}{d x} g(x)_{x=0}=-g(0)/a_{12}$, where $a_{12}$ is the scattering length associated to the inter-species coupling constant $g_{12}$, i.e. $a_{12}=-\frac{2\hbar^2}{mg_{12}}$. The long-distance behaviour of the functions $f(x)$ and $g(x)$ needs to be chosen with care. In fact, while it does not affect the final converged results of our calculations, we find that  an accurate optimization of the functions $f(x)$ and $g(x)$ is crucial for the efficiency of the algorithm. Thus we choose to model the functions $f(x)$ and $g(x)$ with 1D B-splines. The control coefficients of the B-splines are optimized by minimizing the variational energy of the wavefunction. The minimization is accomplished using the stochastic reconfiguration method~\cite{becca_sorella_2017,PhysRevB.61.2599,SR} .
\section{III. Results}
\begin{figure}
\begin{center}
\includegraphics[width=8.5cm]{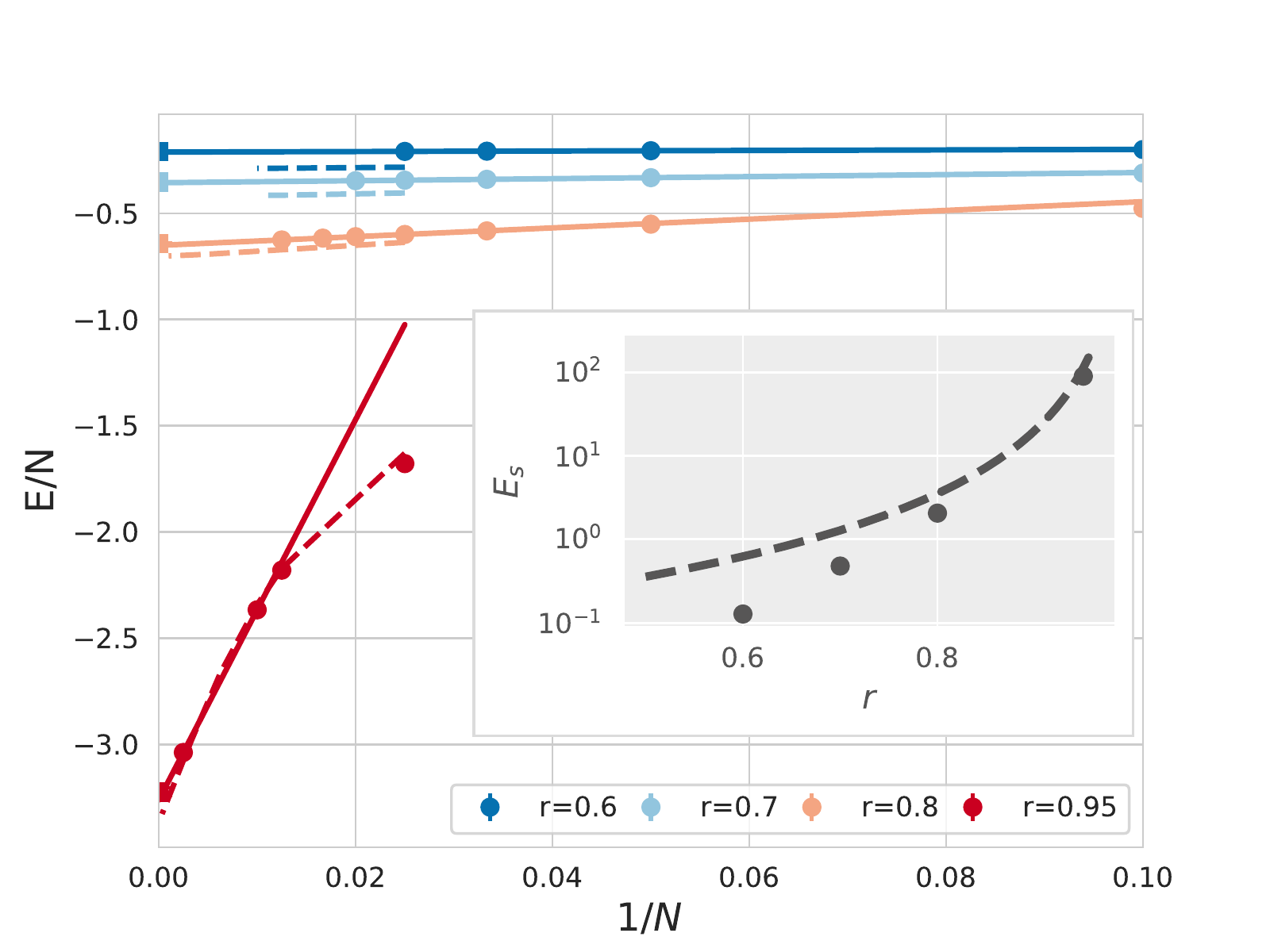}
\caption{Energy per particle, in units of $\hbar^2/(ma^2)$, as a function of the inverse particle number $1/N$ for several values of the ratio $r$  between attraction and repulsion. Solid dots are the results of our DMC simulations and solid lines are linear fits to the DMC data in the limit of large $N$. Solid squares at $1/N=0$ correspond to the bulk energy calculated in Ref.~\cite{parisi-liquid-mixture}. Dashed lines show instead the GGP predictions. The slope of the linear fit, corresponding to the surface energy $E_S$, is shown in the inset together with the GGP result.}
\label{fig1}
\end{center}
\end{figure}

\paragraph{\bf{Surface tension:}}
The first quantity we discuss are the finite-size effects exhibited by the energy per particle. In Fig.~\ref{fig1} we show the energy per particle as a function of the inverse particle number $1/N$ for different ratios $r=|g_{12}|/g$ of the coupling constants. In the limit of infinite particle number, the energy contribution from the surface is negligible compared to the bulk. Moreover, for large $N$, the bulk of the droplet becomes flat and can be described as a uniform liquid at the equilibrium density corresponding to the energy minimum of the uniform mixture. We already performed a systematic study of the properties of the uniform liquid state in Ref.~\cite{parisi-liquid-mixture} and in particular we computed the bulk energy for several values of the ratio $r$. These values are shown in Fig.~\ref{fig1}.

At finite particle number the bulk energy is reduced because of surface effects. In Fig.~\ref{fig1} we plot the energy per particle obtained from our DMC simulations. For large $N$ the liquid-droplet model predicts that the energy should depend linearly on $1/N$. In fact, at large particle numbers, the QMC results are well aligned with the linear fits shown in Fig.~\ref{fig1}. At the ratio $r=0.6$, the linear dependence is satisfied already with very few particles. At larger ratios, closer to the weakly interacting limit, finite size effects increase. For example, at a ratio $r=0.95$, the linear behavior is reached only for $N\gtrsim80$.

In Fig.~\ref{fig1} the QMC results are compared with the GGP predictions (dashed lines in Fig.~\ref{fig1}). The validity of the GGP equation is justified only in the limit in which $|g_{12}|$ is very close to $g$. We find that at $r=0.95$ results are in good agreement, while for lower ratios the GGP equation underestimates the energy of the system. For example, at $r=0.6$ and for $N\approx 40$ the droplet is around $40\%$ less bound than within the GGP theory.

The slope of the linear fit yields the surface energy $E_S$. We plot $E_S$ in the inset of Fig.~\ref{fig1} and compare it with the results of GGP theory. We find that this latter overestimates the surface energy compared with our DMC results, even though good agreement is obtained for large ratios. In conclusion, our results suggest that the GGP equation provides accurate predictions for the surface tension in the weakly interacting limit. Deviations become significant for ratios lower than $r\approx 0.8$.

\begin{figure}
  \begin{center}
   
    \includegraphics[width=8.5cm]{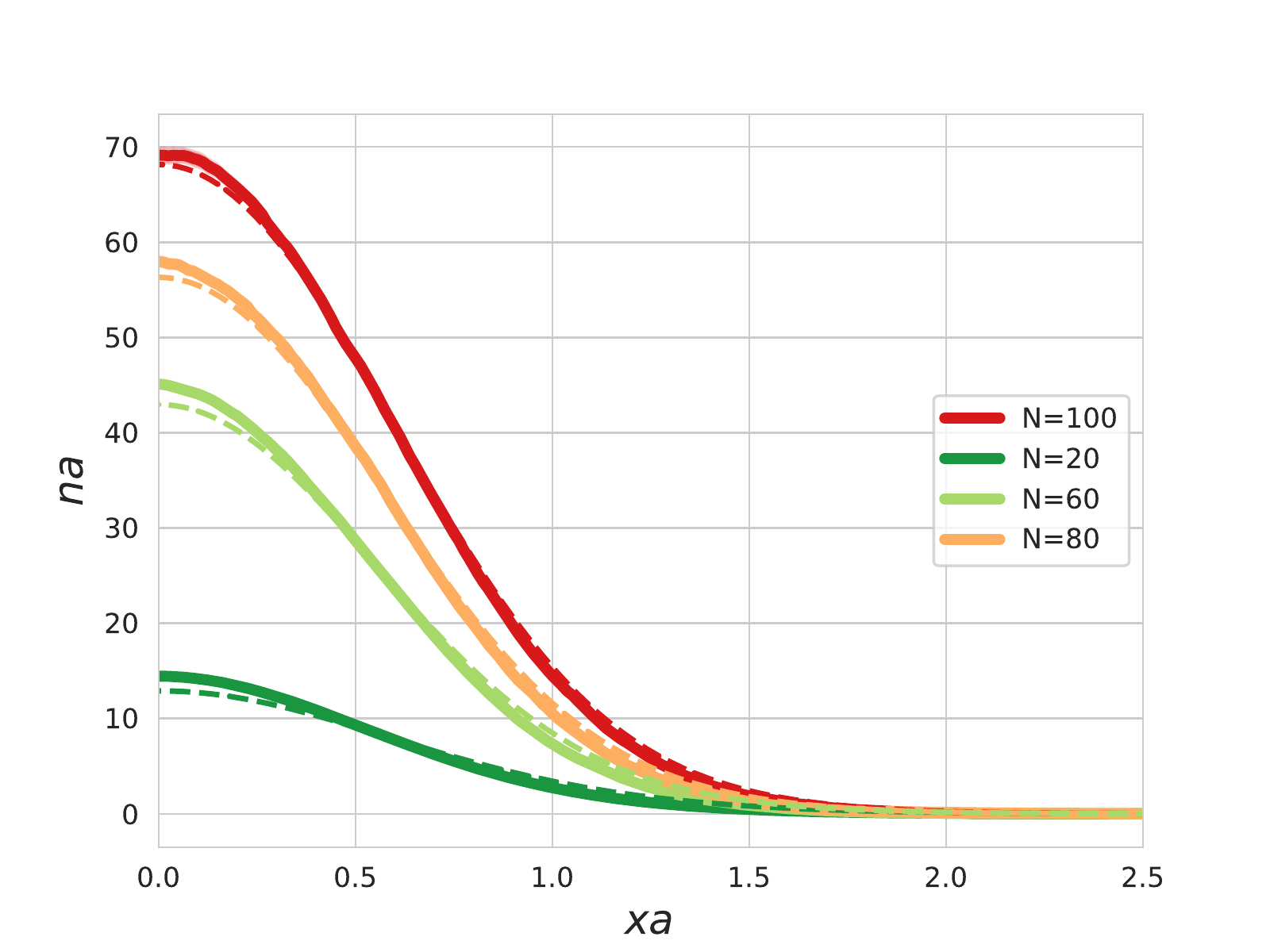}
    \includegraphics[width=8.5cm]{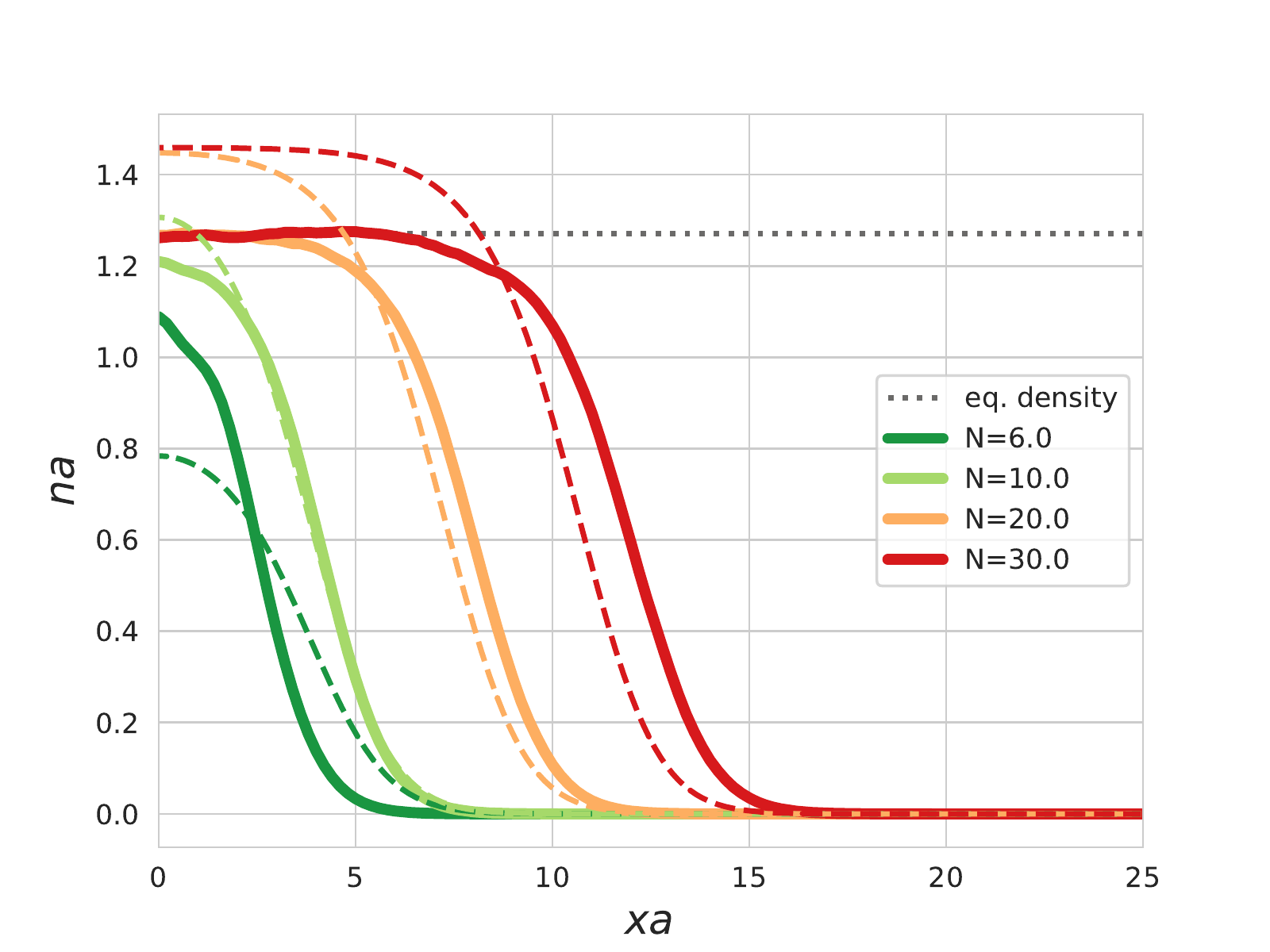}
  
\caption{Density profiles for varying particle numbers $N$ and ratios $r$: $r=0.95$ (top panel) and $r=0.6$ (bottom panel). Dashed lines correspond to the profiles obtained from the GGP equation in the same conditions. The dotted line is the equilibrium density computed for a uniform mixture at $r=0.6$ in Ref.~\cite{parisi-liquid-mixture}.}
\label{fig2}
\end{center}
\end{figure}

\paragraph{\bf{Density profiles:}}
In Fig.~\ref{fig2} we show some typical density profiles, for ratio $r=0.95$ (top panel) and ratio $r=0.6$ (bottom panel), corresponding to different overall normalizations fixed by the number of particles in the droplet. As one increases the particle number the density profile appears higher and broader. In the limit of very large $N$ one expects to reach a regime where the density near the center of the cloud is flat and reaches the equilibrium value of the uniform system. For the ratio $r=0.95$ the saturation density is reached only for very large particle numbers beyond the system size we can reliably simulate. Furthermore, for this ratio, we compare our results with the predictions of GGP theory (dashed lines in top panel of Fig.~\ref{fig2}). We find that the GGP equation predicts slightly lower densities for all particle numbers. The difference between the GGP predictions and our QMC results are just few percent and could also be due to residual beyond mean-field effects, present in homogeneous mixtures, which are not described by the GGP energy functional in Eq.~(\ref{BogoliubovEnergy}).

In the lower panel of Fig.~\ref{fig2} we show the density profiles corresponding to $r=0.6$. For this value of the ratio the role of particle correlations is expected to be more important and one could expect deviations from the GGP predictions. Indeed, we observe striking differences between the density profiles obtained using the two methods. Already for six particles the central peak density is around $20\%$ higher than the GGP prediction. By increasing the atom number, the central density also increases and for $N\approx15$  it saturates to a finite value. For larger $N$ the profile becomes flat at the center of the cloud and, while the radius of the cloud keeps increasing with $N$, the central density no longer depends on the particle number. It is worth noticing that the equilibrium density can be determined by minimizing the energy of a uniform system as it was done in Ref.~\cite{parisi-liquid-mixture}. This density is shown as a dotted line in the bottom panel of Fig.~\ref{fig2}, in perfect agreement with the saturation density of the droplets. The density profiles predicted by the GGP equation follow a qualitatively similar trend, but are quantitatively different. The GGP density profiles saturate at a different density which is around $20\%$ higher. This is compatible with the results for uniform systems~\cite{parisi-liquid-mixture}.

\begin{figure}
\begin{center}
\includegraphics[width=8.5cm]{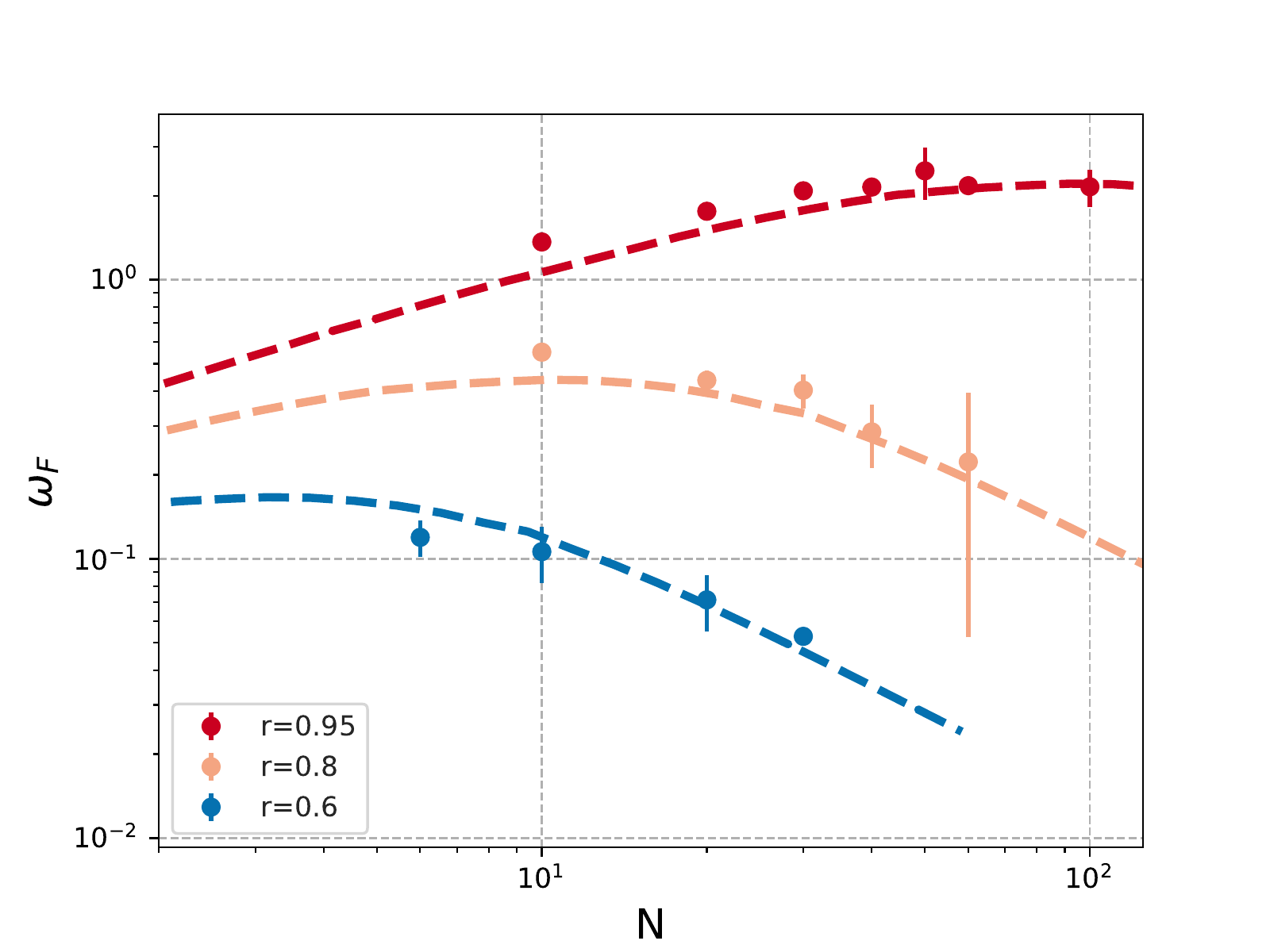}
\caption{Frequency of the breathing mode, in units of $\hbar/(ma^2)$, for several values of the particle number $N$ and the ratio $r$. Solid dots correspond to DMC results, dashed lines to GGP predictions.}
\label{fig3}
\end{center}
\end{figure}

\paragraph{\bf{Breathing mode:}}
We provide now an estimate of the frequency of the breathing mode for different particle numbers and ratios of coupling strengths. We make use of Eq.~(\ref{breathingModeSumRule}) and we compute the average size of the cloud and its mean square fluctuations. In the case of the GGP equation, we solve the time dependent equation starting from a configuration out of equilibrium and we determine the frequency of the oscillations in the cloud size. To obtain the initial condition we first propagate the GGP equation in imaginary time to find the ground-state wavefunction of the system. We then multiply the ground-state wavefunction by a factor $e^{-i \delta x^2/a^2}$, where $x$ is the spatial coordinate and $\delta$ is a dimensionless parameter which controls the strength of the perturbation. We make sure to choose $\delta$ sufficiently small in order to excite just the breathing mode. Our results are shown in Fig.~\ref{fig3}. Overall we find a very good agreement between DMC and GGP results. Surprisingly, the agreement is very good for all values of the ratio $r$ especially at large particle numbers.

For a fixed value of the ratio between attraction and repulsion, the GGP theory predicts that the frequency of the breathing mode first increases by increasing the atom number, reaches a maximum at a certain critical $N$ and then decreases approaching zero for very large particle numbers. Correspondingly, the nature of the breathing mode changes from a surface mode, driven by surface tension effects, to a bulk density mode driven by the bulk compressibility. In the limit of large $N$ the excitation energy decreases due to its scaling with the inverse size of the droplet.

At the ratio $r=0.95$ and small atom numbers, we find small deviations compared to the GGP equation, which seems to underestimate the frequency of the breathing mode. Such deviations, however, disappear when the particle number increases and our results agree with GGP predictions within statistical errors. The same occurs for lower ratios, $r=0.8$ and $r=0.6$. This result is surprising because of the clear differences with GGP theory which emerged at these values of $r$ in the surface tension $E_S$ (see Fig.~\ref{fig1}) and in the density profiles (see bottom panel of Fig.~\ref{fig2}). A possible explanation is that the two deviations compensate and the frequency of the breathing mode is well reproduced by GGP theory.

\section{IV. Conclusions}
In conclusion we have carried out a systematic QMC study of the properties of quantum droplets in 1D mixtures and a quantitative comparison with GGP theory for a number of relevant physical quantities such as the surface tension, the density profiles and the frequency of the breathing collective mode. Overall we find a good agreement between our exact results and the predictions of GGP theory. Deviations are small, even when the ratio between coupling constants is significantly below the value $r=1$ where the approximate theory is expected to hold. Our results will also be useful to guide future experiments on low-dimensional quantum droplets.

{\it Acknowledgements:} We thank G.E. Astrakharchik for useful discussions. This work was supported by the QUIC grant of the Horizon 2020 FET program and by Provincia Autonoma di Trento. 

\bibliographystyle{unsrt}

\bibliography{articleDroplet}

\end{document}